\documentclass[aps,11pt,onecolumn,preprintnumbers,amsmath,amssymb,superscriptaddress]{revtex4}

\usepackage[english]{babel}
\usepackage{graphicx}
\usepackage{color}
\usepackage{epstopdf}

\begin{document}

\newcommand{\unit}[1]{\:\mathrm{#1}}            % write units
\newcommand{\To}{\mathrm{T_0}}

\newcommand{\Tp}{\mathrm{T_+}}
\newcommand{\Tm}{\mathrm{T_-}}
\newcommand{\EST}{E_{\mathrm{ST}}}
\newcommand{\Rp}{\mathrm{R_{+}}}
\newcommand{\Rm}{\mathrm{R_{-}}}
\newcommand{\Rpp}{\mathrm{R_{++}}}
\newcommand{\Rmm}{\mathrm{R_{--}}}
\newcommand{\ddensity}[2]{\rho_{#1\,#2,#1\,#2}} % for diagonal density matrix elements
\newcommand{\ket}[1]{\left| #1 \right>} % for Dirac bras
\newcommand{\bra}[1]{\left< #1 \right|} % for Dirac kets

\title{Zeeman splitting via spin-valley-layer coupling in bilayer MoTe$_2$}
\author{Chongyun Jiang}
\affiliation{Division of Physics and Applied Physics, School of Physical and Mathematical Sciences, Nanyang Technological University, Singapore 637371, Singapore}
\author{Fucai Liu}
\affiliation{Center for Programmable Materials, School of Materials Science \& Engineering, Nanyang Technological University, 50 Nanyang Avenue, Singapore 639798, Singapore.}
\author{Jorge Cuadra}
\affiliation{Division of Physics and Applied Physics, School of Physical and Mathematical Sciences, Nanyang Technological University, Singapore 637371,
	Singapore}
\author{Zumeng Huang}
\affiliation{Division of Physics and Applied Physics, School of Physical and Mathematical Sciences, Nanyang Technological University, Singapore 637371, Singapore}
\author{Ke Li}
\affiliation{Division of Physics and Applied Physics, School of Physical and Mathematical Sciences, Nanyang Technological University, Singapore 637371, Singapore}
\author{Ajit Srivastava}
\affiliation{Department of Physics, Emory University, Atlanta, Georgia 30322, USA}
\author{Zheng Liu}
\affiliation{Center for Programmable Materials, School of Materials Science \& Engineering, Nanyang Technological University, 50 Nanyang Avenue, Singapore 639798, Singapore.}
\author{Wei-bo Gao}
\affiliation{Division of Physics and Applied Physics, School of Physical and Mathematical Sciences, Nanyang Technological University, Singapore 637371, Singapore}

%\pacs{42.50.Ex, 42.50.Lc, 78.67.Hc}

\begin{abstract}

%\textbf{The coupling of spin and valley degrees of freedom lies at the heart of emerging field of valleytronics based on transition metal dichalcogenides (TMDs). Unlike monolayer TMDs, bilayers possess an additional degree of freedom viz., the layer pseudospin which makes the spin-orbital coupling not only relate to spin and valley, but also different layers. Recently, a magnetic control of the valleys in monolayer TMDs was demonstrated in a valley analog of the Zeeman effect. Here, we find that the Zeeman splitting of the circularly polarized components of the optical  emission persists in bilayer TMDs and the degree of circularly polarized photoluminescence can be tuned with magnetic field from $-50\%$ to $50\%$. Although the valley degeneracy is not lifted, the locking of circular polarization with spin together with unequal magnetic moments of electrons and holes results in a Zeeman effect in each valley. This stems from the coupling of spin, valley and layer pseudospins, more specifically, the opposite sign of the magnetic moment in the same valley for different spin-layers. The measured g-factor of the splitting provides an estimate of the interlayer coupling. Our results demonstrate spin-valley-layer coupling and together with previous electric field control, pave the way for valleytronics based on bilayer TMDs.}

\textbf{Atomically thin transition metal dichalcogenides (TMDs) possess coupling of spin and valley degrees of freedom, making them promising for spin-valleytronics~\cite{XiaoLiuEtAl2012, Xu14, Geim13}. %Unlike monolayer TMDs, bilayers possess an additional degree of freedom viz., the layer pseudospin which is coupled to spin and valley and can be exploited for spin-valley quantum gates based on magnetoelectric effects~\cite{}.
ln monolayer TMDs, the emission helicity is locked to the valleys as a consequence of spin-orbit coupling and inversion symmetry breaking, leading to a valley analog of Zeeman effect in presence of out-of-plane magnetic field~\cite{MacNeillHeikesEtAl2015,SrivastavaSidlerEtAl2015a,AivazianGongEtAl2015,LiLudwigEtAl2014,StierMcCrearyEtAl2016,AroraSchmidtEtAl2016}. As inversion symmetry is recovered in bilayers, the emission helicity should no longer be locked to the valleys. Here we show that Zeeman splitting, however, still persists in bilayers, as a result of an additional degree of freedom viz., the layer pseudospin and spin-valley-layer locking~\cite{Xu14}. In contrast to monolayer, Zeeman splitting here exists without lifting valley degeneracy. The degree of circularly polarized photoluminescence can be tuned with magnetic field from $-27\%$ to $27\%$. Our results demonstrate the control of degree of freedom in bilayer with magnetic field, which, together with previous electric field control, %in the toolbox, 
makes bilayer a promising platform for spin-valley quantum gates based on magnetoelectric effects~\cite{GongLiuEtAl2013}. }

\end{abstract}

\maketitle

%\subsection{Introduction}

In monolayer group VI transition metal dichalcogenides (TMDs) such as MoS$_2$ and WSe$_2$, spatial inversion symmetry is broken, leading to finite but opposite Berry curvature and magnetic moment in the two valleys~\cite{XiaoLiuEtAl2012}. Due to strong spin-orbit coupling, spin and valley degrees of freedom are locked which results in a series of exotic valley effects, such as valley Hall effect~\cite{LiEtAl2013,MakMcGillEtAl2014}, valley optical selection rule~\cite{MaK12,Zeng12,Cao12,Jones13} and valley Zeeman splitting \cite{MacNeillHeikesEtAl2015,SrivastavaSidlerEtAl2015a,AivazianGongEtAl2015,LiLudwigEtAl2014,StierMcCrearyEtAl2016,AroraSchmidtEtAl2016}. In bilayer TMDs, the layers are rotated by 180$^\circ$ with respect to each other, resulting in the recovery of spatial inversion symmetry. It is then natural to ask whether the above-mentioned valley phenomena still exist in bilayer TMDs. Interestingly, when the interlayer coupling is much smaller than the spin-orbit interaction, bilayer TMD can be regarded as two decoupled monolayers with the layer pseudospin leading to a spin-valley-layer coupling. This forms a potential platform for spin-valley quantum gates with magnetic and electric control \cite{GongLiuEtAl2013}. To this end, spin-layer locking induced valley Hall effect \cite{JieunLee16}, valley optical selection rule \cite{BairenZhu14} and electric control \cite{JonesYuEtAl2014} have been experimentally demonstrated. In this work, we demonstrate that magnetic field can be another addition in the toolbox to explore the rich interplay between real spin and valley, layer pseudospins in bilayer TMDs. Specifically, we find that the Zeeman splitting of the two circularly polarized components of optical emission persists in bilayer TMDs regardless of the recovery of inversion symmetry due to spin-valley-layer locking. The magnetic field control, together with electric control as demonstrated previously, paves the ways for quantum manipulation of spin, valley, and layer degrees of freedom in bilayer TMDs \cite{GongLiuEtAl2013}.  

We perform our experiments on MoTe$_2$, which is a layered material with honeycomb lattice. With decreasing number of layers, the indirect gap of bulk MoTe$_2$ turns into direct gap at valleys (K, K'). Berry curvature and orbital magnetic moments can be studied through the polarization of the emitted light through valley optical selection rule.  MoTe$_2$ is a semiconductor with a relatively smaller band gap and its photoluminescence (PL) emission is in the near infrared range around $\sim$1.1 eV. A reversible structural phase transition between hexagonal and stable monoclinic has been reported in bulk single-crystalline MoTe$_2$ \cite{KeumChoEtAl2015} and a semiconductor-to-metal electronic phase transition has been demonstrated by thinning down bulk MoTe$_2$ or straining the MoTe$_2$ thin films \cite{SongKeumEtAl2016}. These features make MoTe$_2$ a flexible material suitable for valley-based optoelectronic applications.

%MoTe2 acts an ideal platform for bilayer physics since the emission in MoTe2 bilayers are in the same order of monolayer, as compared to rapid decreasing of the emission intensity with increased layer number in other TMDCs such as MoS2, WSe2.

\begin{figure}
\includegraphics[scale=0.5]{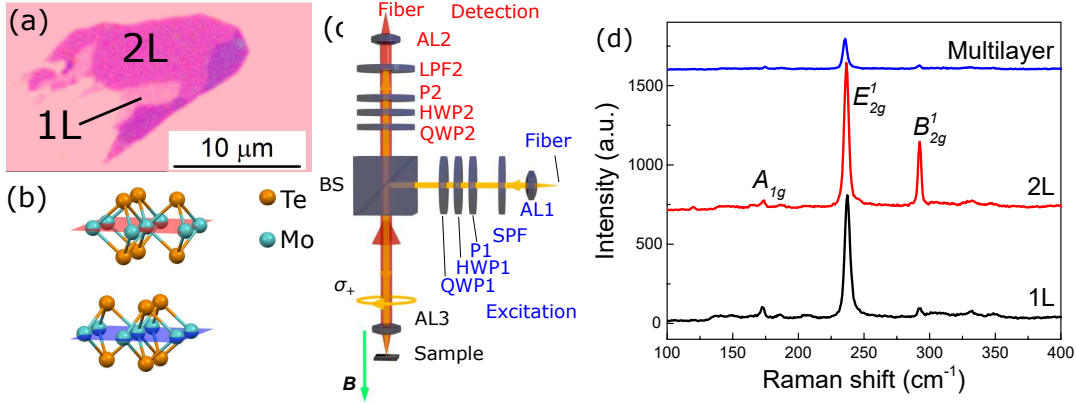}
\caption{ \textbf{Sample characterization.} \textbf{a}, Optical microscope image of the MoTe$_2$ monolayer and bilayer. \textbf{b}, Crystal structure of a bilayer MoTe$_2$. The two layers are rotated in plane by $180^\circ$ relative to each other. \textbf{c}, Optical setup for the polarization-resolved PL spectroscopy. The optical components are: achromatic lenses (AL1-3), polarizers (P1 and P2), half wave plates (HWP1 and HWP2), quarter wave plates (QWP1 and QWP2), a 1050 nm short pass filter (SPF), a 1050 nm long pass filter (LPF) and a beam splitter (BS). The sample is placed in a helium bath cryostat with an out of plane magnetic field in a Faraday geometry. The green arrow shows a negative magnetic field. \textbf{d}, Raman spectroscopy of the MoTe$_2$ monolayer, bilayer and multilayer. $A_{1g}$, $B_{2g}^1$ and $E_{2g}^1$ represent different modes in Raman spectrum.}\label{fig:1}
\end{figure}

The optical image of the sample used in the study is illustrated in Figure \ref{fig:1}a, where the monolayer (1L) and bilayer (2L) can be easily identified by their optical contrasts.
The flakes are mechanically exfoliated using adhesive tapes and then transferred onto silicon wafers with a 300 nm thick thermally grown SiO$_2$. The as-prepared sample are kept under vacuum to prevent oxidation and deliquesce. The crystal structure of a bilayer AB-stacked MoTe$_2$ is shown in Figure \ref{fig:1}(b). The upper layer is rotated by 180$^\circ$ with respect to the lower layer such that the inversion symmetry is recovered as compared to monolayer.  Layered MoTe$_2$ has an exciton energy of $\sim$ 1.1 eV, which can be experimentally extracted by the photoluminescence spectroscopy.  We utilize a home-made fiber-based confocal microscope setup for the micro-PL experiments (see Figure \ref{fig:1}(c)). We show the details of our experimental setup in Methods section.  The excitation and collection polarizations are controlled by a series of polarizers and quarter-wave plates. Below, we refer to co-polarization (cross-polarization) when the quarter-wave plates are configured for the same (opposite) handedness.

%
%We cool down the sample to 4 K in a Quantum Design Physical Properties Measurement System (PPMS) with a superconducting magnet which can generate a magnetic field up to $\pm 9$ T. A tunable diode laser at wavelength of 1.19 eV (1140 nm) is used as the excitation source.  The excitation laser is spectrally cleaned after passing through a grating and a short pass filter giving a bandpass of 0.3 nm. The laser beam is then focused to a spot size $\sim \mu m$ on the sample with an objective (NA = 0.75).  

To further confirm the number of layers in our sample, we perform Raman spectroscopy of the monolayer, bilayer and multilayers at room temperature as shown in Figure \ref{fig:1}d.  The $B_{2g}^1$ mode of the 2L is strong with a Raman shift of 292.4 cm$^{-1}$ while that of the 1L and multilayer is very weak. The in-plane mode $E_{2g}^1$ (out-of-plane mode $A_{1g}$) exhibits downshift (upshift) in energy as the number of layer increases. The results agree well with the previous report \cite{PradhanRhodesEtAl2014,YamamotoWangEtAl2014}, confirming the number of layers of the investigated sample.

%The samples characterization and our experimental schematic setup is shwon in Fig. 1. Optical microscope image of the mechanically exfoliated sample is shown in Figure 1a, where monolayer and bilayer can be identified by optical contrast.  

% The excitation power is $\sim$400 $\mu$W over a focused laser spot area of $\sim$ 1 $\mu$m in diameter. Linear polarizers and achromatic quarter-wave plates are adopted on the excitation and detection arm of the confocal microscope for polarization-resolved PL spectroscopy acquisition.  The axes of linear polarizers are aligned horizontally for initializing and analyzing the polarization state of the incoming and outgoing light. The quarter-wave plate (QWP1) on the excitation arm converts linearly polarization into right- ($\sigma_+$) or left- ($\sigma_-$) circular polarization.  The quarter-wave plate (QWP2) on the detection arm with the linear polarizer (P2) can extract circularly polarized PL emission from the sample.Figure \ref{fig:1}(d) shows the Raman spectroscopy of the 1L and 2L, and a multilayer in the vicinity for reference at room temperature. The excitation energy is 2.331 eV (532 nm).  The Raman shift for the in-plane $E_{2g}^1$ mode are 237.3 cm$^{-1}$, 236.5 cm$^{-1}$ and 235.8 cm$^{-1}$ for the 1L, 2L and multilayer MoTe2. The Raman shift for the out-of-plane $A_{1g}$ mode are 173.1 cm$^{-1}$, 173.9 cm$^{-1}$ and 174.6 cm$^{-1}$.

%For all the orientations of the wave plates we have to compensate the phase shift of the beam splitter (BS). Therefore, the rotations of the fast axes differ from $\pm 45^\circ{}$.

After sample characterization, we next demonstrate the observed Zeeman splitting in PL spectrum and PL polarization control by magnetic field. Following these, we will discuss the origin of our observations. Figure 2a and 2b show the polarization-resolved PL spectra of monolayer and bilayer MoTe$_2$ under external magnetic field of -8 T, 0 T and +8 T perpendicular to the sample plane at 4 K. 
The monolayer shows single emission at 1.162 eV (Peak $A_1$ in Fig. 2a) and the bilayer sample shows multiple emissions at 1.153 eV and 1.135 eV (Peak $A_2$ and Peak $B_2$ in Fig. 2b). Here, peak $A_1$ and $A_2$ are attributed to the optical transition of the neutral excitons in the 1L and 2L respectively. Peak $B_2$ corresponds to the transition of charged exciton (trion) state in the 2L \cite{YangLueEtAl2015}. From the figure, we can make two main observations: First, at zero magnetic field the wavelengths of PL emission are at the same position for  $\sigma_+$ and $\sigma_-$ detection. At -8 T (+8 T), however, the position of the peaks blueshift (redshift) for $\sigma_+$ ($\sigma_-$) detection, which indicates an energy splitting. Second, the magnitudes of the peaks for $\sigma_+$ and $\sigma_-$ detection under magnetic field also differ, which manifests itself as magnetic field dependent PL polarization. Here the degree of the PL polarization can be defined as 
$\eta_{PL}=\frac{I_{\sigma_+}-I_{\sigma_-}}{I_{\sigma_+}+I_{\sigma_-}}$, where $I_{\sigma_+}$ ($I_{\sigma_-}$) is the intensity of the PL emission in $\sigma_+$-out ($\sigma_-$-out) configurations. 

%The main finding of this letter is that magnetic field dependent Zeeman splitting and PL polarization do not only exist in monolayer, but also in bilayers even with the recovery of spatial inversion symmetry. 

\begin{figure}
\includegraphics[scale=0.5]{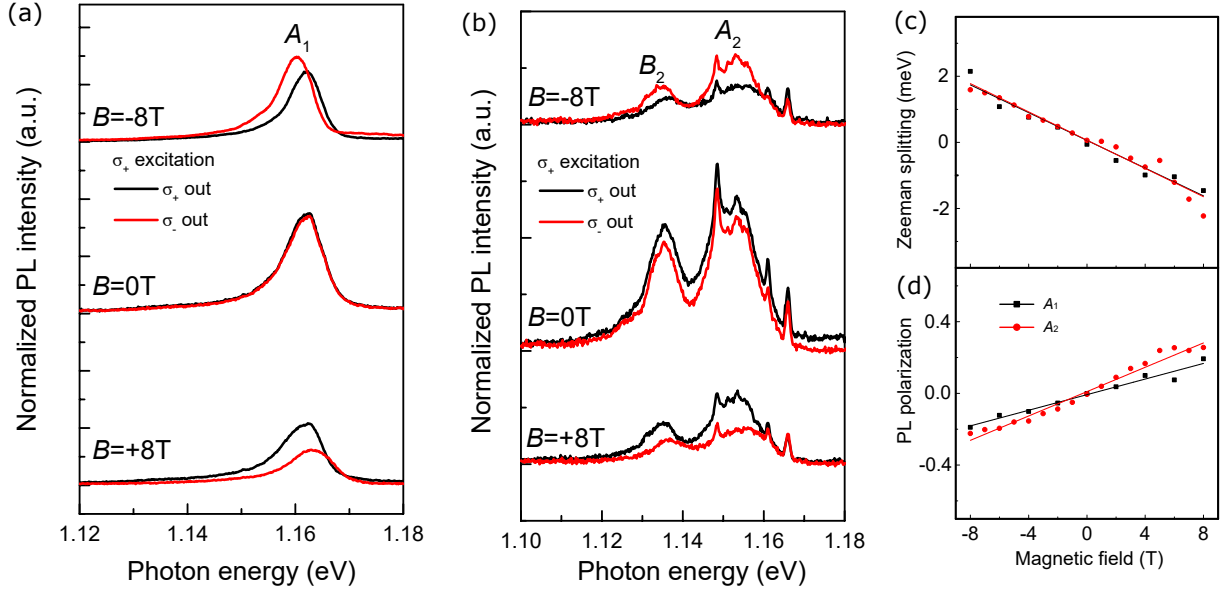}
\caption{ \textbf{Zeeman splitting comparison for monolayer and bilayer.} Polarization-resolved PL spectra taken at -8 T, 0 T and +8 T at 4 K from (a) monolayer and (b) bilayer MoTe$_2$. Positive magnetic field is in the direction against the propagation of the incoming excitation. The excitations are set to $\sigma_+$ circular polarization. The detection polarization is configured to $\sigma_+$ (black solid lines) and $\sigma_-$ (red solid lines) circular polarization. The PL intensity of the bilayer is one order of magnitude smaller than that in the monolayer due to higher symmetry of the bilayer.  Here the PL spectra is normalized by emission maximum for each particular magnetic field and offset for better visualization. Sharp peaks in bilayer PL spectrum are from Raman peaks.  (c) Zeeman splitting and (d) PL polarization versus magnetic field comparison for monolayer and bilayer MoTe$_2$ at 4 K.} \label{fig:2}
\end{figure}

To further illustrate these results, spectrum splitting and PL polarization for the neutral excitons are quantitatively depicted in Fig. 2c and Fig. 2d, respectively. 
The Zeeman splitting of an optical transition is fit with $\Delta E=g\mu_B B$, where $g$ is a $g$-factor that is associated with the magnetic moment in the system, $\mu_B$ is the Bohr magneton and $B$ is the magnetic field. The optical transition of the exciton state (Peak $A_1$ of 1L and $A_2$ of 2L) depends linearly on the magnetic field, with slopes of -211$\pm14$ $\mu$eV/T and -214$\pm10$ $\mu$eV/T, corresponding to $g(A_1)=3.65\pm 0.24$ and $g(A_2)=3.70\pm 0.17$. Similar to Zeeman splitting, here the PL polarization of the 1L and 2L also depends linearly on the magnetic field.  Here we focus on the emission polarization only and the average PL polarization for $\sigma_+$ and $\sigma_-$ excitation is shown. We fit the relationship of the PL polarization $\eta_{PL}$ and the magnetic field $B$ with $\eta_{PL}=\beta B$, where  $\beta$ is a coefficient with (2.18$\pm 0.15)\times 10^{-2}$ T$^{-1}$ for Peak $A_1$ of the 1L and (3.41$\pm 0.12)\times 10^{-2}$ T$^{-1}$  for Peak $A_2$ of the 2L. Full spectrum for 1L and 2L, the fit result of the trion peak $B_2$ and local exciton  are shown in the Supplementary information. From supplementary Figure S4, we see that there is a finite valley polarization for 2L; however, the predominant part of the PL polarization is not associated with the polarization state of the excitation. This points to a fast valley mixing during the creation of low energy excitons even under quasi-resonant excitation.

The result of monolayer Zeeman splitting is not new and such energy splitting was measured before and attributed to the lifting of the valley degeneracy in the band structure due to the breaking of the time reversal symmetry in the presence of a magnetic field, so called valley Zeeman splitting, or valley splitting for short \cite{MacNeillHeikesEtAl2015,SrivastavaSidlerEtAl2015a,AivazianGongEtAl2015,LiLudwigEtAl2014}. The main observation here is that such Zeeman splitting still persists in bilayer, which comes as an unexpected result due to the inversion symmetric nature of bilayers. Below, we focus on the physical origin of such splittings.

\begin{figure}
\includegraphics[scale=0.2]{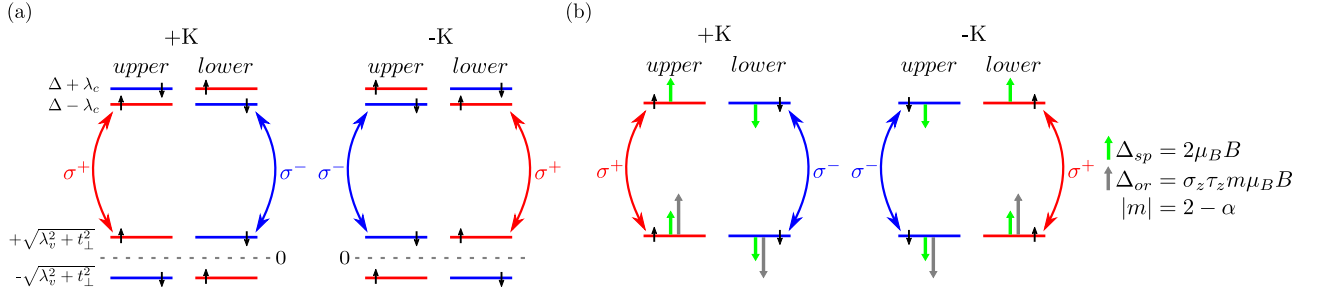}
\caption{\textbf{Origin for Zeeman splitting in bilayer.} (a) Single particle energy states at $\pm K$-points of bilayer TMD at zero magnetic field and in presence of interlayer hopping $t_\perp$. The spin-valley-layer locking results in optical selection rules such that in both valleys, the spin degree of freedom is locked to the emission helicity. (c) Schematic diagram of the Zeeman splitting in bilayer MoTe$_2$ under positive magnetic field at $\pm K$-points. Red (blue) transition indicates the PL emission with photon energy $E_+$ ($E_-$) and circular polarization $\sigma_+$ ($\sigma_-$). The upper and the lower layer in the same valley have opposite spin. Green and grey arrows show the magnetic moment contributions of the spin and atomic orbital to the Zeeman splitting, respectively. In presence of a positive magnetic field, the spin contributions cancel while the intracellular or atomic orbital contribution leads to $E_+ < E_-$ in both the valleys.}
\end{figure}

In monolayer TMDs, the real spin and the valley pseudospin are effectively coupled by spin-orbital coupling and broken inversion symmetry \cite{XiaoLiuEtAl2012}. Bilayer TMDs possess another degree of freedom, viz., layer pseudospin~\cite{GongLiuEtAl2013}.  In bilayer, the Hamiltonian at $\pm K$ points can be expressed in a two-band approximation as $H_c = \Delta + \lambda_c \tau_z s_z \sigma^c_z$ for the conduction band and $H_v = -\lambda_v  \tau_z s_z \sigma^v_z + t_\perp \sigma^v_x$ for the valence band, where $\Delta$ is the band gap, $\lambda_c$ ($\lambda_v$) denotes the spin-orbit coupling of conduction (valence) band and $t_\perp$ is the interlayer coupling of the layers. The strong coupling between the valley ($\tau_z$) and layer ($\sigma^{c,v}_z$) pseudospin, and the real spin ($s_z$) is a distinguishing feature of bilayers. The layer Pauli operators $\sigma^c_z$ ($\sigma^v_{z}$) are expressed in the basis of $d_{z^2}$ ($d_{x^2 - y^2} \pm i d_{2xy}$). The interlayer hopping, $t_\perp$, vanishes for conduction band due to the symmetry of $d_{z^2}$ orbitals. When the spin-orbit coupling strength $\lambda_v$ is much larger than the interlayer hopping amplitude, holes are primarily confined to either upper or lower layer, which can be labeled with layer pseudospin up $|u\rangle$ or down $|l\rangle$. 

Figure 3a depicts the energy level diagram at zero magnetic field emphasizing the spin-valley-layer locking in bilayer. At a given energy in a given valley, different layers carry opposite spins. The lowest energy single-particle optical transitions giving rise to excitonic resonance for different valley, layer and spins are also shown in Fig.~3a. As the spin is conserved in the optical transition (singlet exciton), spin-valley-layer locking leads to emission helicity being locked with the spin degree of freedom in both valleys. Upon diagonalizing $H_v$, the hole energies shift from $\pm \lambda_v$ to $\pm \sqrt{\lambda_v^2 + t^2_\perp}$ and the new eigenstates are admixtures of $d^u \pm i d^u$ and $d^l \mp i d^l$ orbitals. Unlike the case of monolayer where helicity of emission is tied to the valley degree of freedom, optical transitions of either helicity are present in both valleys for bilayers. In the absence of magnetic field, all four optical transition depicted in Fig.~3a are degenerate.   
%
%In bilayer, near the $\pm K$ point, the Hamiltonian can be written as $H_v=-\varepsilon-\lambda\tau_z\sigma_zs_z+t_\perp\sigma_x$, where $\varepsilon$ is energy dispersion, .   First Brillouin zone of bilayer is drawn in Figure. 3a. At the same valley, spin is flipped for different layers. 
%
%Optical transitions for different valley and layer are illustrated in Fig. 3b. All four transitions in $|K, u\rangle$,  $|K, l\rangle$, $|-K, u\rangle$, $|-K, l\rangle$ have the same energy for zero magnetic field, leading to spectrum degeneracy. 

When an out of plane $B$ field is applied, conduction and valence band energy will be shifted, in accordance with the respective magnetic moments as shown in Fig.~3b. The conduction band states have contribution only from the  spin as $d_z^2$ orbitals do not carry any magnetic moment whereas the valence band states have orbital magnetic moment (intracellular contribution) stemming from $d \pm i d$ orbitals in addition to the spin contribution. The intercellular contribution which plays a role in monolayer TMDs vanishes in bilayers due to restoration of spatial inversion symmetry \cite{XiaoEtAl2007}. The spin Zeeman shift can be written as $\varDelta_s=2s_z\mu_BB$. Since $\varDelta_s$ has the same value for conduction and valence band, it thus does not contribute to the net energy shift. Thus, only intracellular contribution which, differs for the two bands, causes a measurable shift in the optical transition energies.

In the limit of negligible interlayer coupling, the valence band is mainly comprised of $d \pm id$-orbitals with $m=\pm2$ while the conduction band has $m = 0$. This intracellular contribution leads to a valley Zeeman splitting with a $g$-factor of 4 in monolayer TMDs. The bilayer case is in stark contrast with this as can be seen from Fig.~3b - although the valley degeneracy is not lifted, each valley experiences a splitting of emission helicity ($\sigma^+ / \sigma^-$) due to intracellular contribution. In other words, while there is a lifting of degeneracy of $\sigma^+ / \sigma^-$ emission in bilayers in the presence of $B$ field, it does not imply a valley Zeeman splitting as the emission helicity is no longer tied to the valley degree of freedom. Instead, the helicity of emission is tied to the spin degree of freedom. A $g$-factor of 4 is thus expected for bilayer Zeeman splitting as well, however, due to finite interlayer hopping, the valence band states are no longer purely $d + i d$ or $d - id$ but an admixture of the two. 

The exact eigenstates of $H_v$ at K-valley ($\tau_z = 1$) and spin up ($s_z = 1$) are given by $u_+ = (\mathrm{cos}~\theta/2, \mathrm{sin}~\theta/2)^T$ and $u_- = (\mathrm{sin}~\theta/2, \mathrm{cos}~\theta/2)^T$ in the basis of where $\mathrm{cos}~\theta = \frac{\lambda} {\sqrt{\lambda^2 + t_\perp^2}}$. Thus, the magnetic moment of valence band states reduces from m = $\pm 2$ to $\tilde{m} = (\pm 2)\mathrm{cos}^2\theta/2 + (\mp 2) \mathrm{sin}^2\theta/2 = \pm 2 \mathrm{cos}~\theta = \pm 2 \frac{\lambda} {\sqrt{\lambda^2 + t_\perp^2}}$. This would imply a Zeeman splitting $g$-factor of $\frac{4\lambda} {\sqrt{\lambda^2 + t_\perp^2}}$ which, for our experimentally measured value of $g$ = 3.7 gives a value of $t_\perp/ \lambda$ = 0.41. Assuming a spin-orbit coupling $\lambda_v$ of $\sim$ 125 meV~\cite{RuppertAslanEtAl2014}, we obtain a interlayer coupling of $\sim$50 meV which is in good agreement with the calculations for similar bilayer TMDs~\cite{GongLiuEtAl2013}.

\begin{figure}
\includegraphics[scale=0.27]{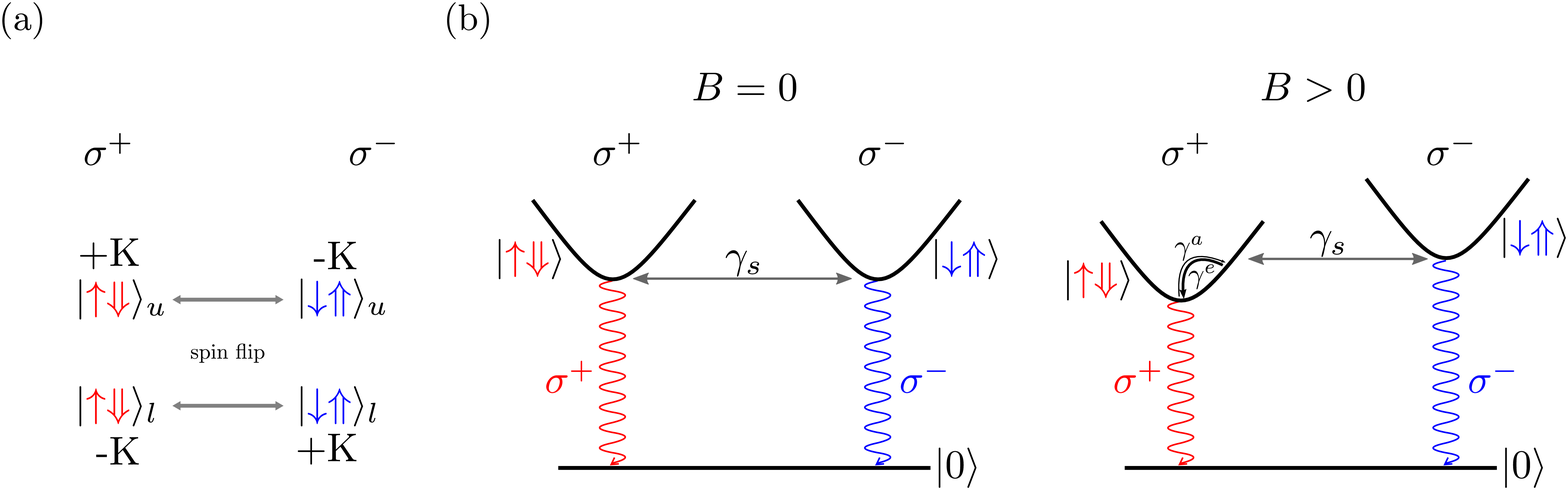}
\caption{\textbf{Origin for PL polarization in bilayer.} (a) The requirement for the switching of light helicity can be thought of as interconversion between circularly polarized excitons, $\sigma_{ex}^+$ and $\sigma_{ex}^-$. The emission helicity is tied to the spin of the electron in the exciton which is formed out of electron-hole pairs in both valleys/layers. The hole spin is labeled by thick vertical arrows and in our convention have spin which is opposite to that of the electron. Due to spin-valley-layer locking, both spin and valley need to be flipped for the switching of helicity, in the absence of interlayer hopping. (b) At positive magnetic field, the selective conversion of $\sigma^-_{ex}$ to $\sigma_{ex}^+$ is energetically favorable due to Zeeman splitting (see main text for details).}
\end{figure}

Finally, we discuss the magnetic field dependence of $\eta_{PL}$ shown in Fig.~2d. As the PL polarization is primarily independent of the excitation polarization, we can conclude that there is fast spin relaxation which leads to creation of both $\sigma^+$ and $\sigma^-$ excitons upon excitation. At zero field, conversion of $\sigma^+$ to $\sigma^-$ and vice versa is equally likely leading to emission from both helicities, as dictated by time-reversal symmetry. At finite $B$-field, the emission intensity of the lower energy peak is always larger. This is true even when the polarity of the $B$-field is reversed implying that the higher energy exciton is transformed into the lower energy exciton with the opposite emission helicity on a timescale which is comparable to exciton lifetime. 

If we assume that the interlayer coupling is suppressed due to large spin-orbit coupling, the conversion of a $\sigma^+$-exciton to a $\sigma^-$-exciton and vice versa requires flipping of both spin and valley degrees of freedom, as shown in Fig.~4a. 
 %The flipping of valley requires a large change in linear momentum $\Delta k$ $\sim$ $2\pi/a$ where $a$ is the lattice constant and should be suppressed in relatively clean samples. However, as shown Fig.~4a, a $\sigma^+$-exciton is made of electronic states with the same spin states but occupying both valleys/layers. Hence, a simultaneous flip of spins in both layers flips the emission helicity of the exciton with no net linear momentum transfer required ($\Delta k = 0$). 
 The spin angular momentum required for such a process is possibly provided during scattering with residual charge carriers present in the sample due to accidental doping. While at zero $B$-field, such a spin flip induced conversion of exciton helicity can occur in both directions, at finite $B$-field, conversion to the lower energy exciton is energetically favorable. To explain the dependence of $\eta_{PL}$ with $B$-field, we assume that the spin flip process is energy conserving while the energy relaxation via phonons primarily conserves spin. Although spin flip via phonon is possible in presence of spin-orbit coupling, it is usually slower than spin-conserving processes~\cite{WangLiuEtAl2013}.  As shown in Fig.~4b, at finite field, spin flip can happen from the higher energy exciton to the excited states of lower energy exciton band at the same energy, which then relax to the lowest energy states by phonons. The reverse process must first involve phonon absorption followed by spin flip due the absence of opposite spin states for the lowest energy exciton. As the phonon absorption is suppressed by the Boltzmann factor, $\mathrm{exp}\left(-\Delta_B/k_B T\right)$ for a Zeeman splitting of $\Delta_B$, the intensity of PL from the lowest energy exciton is dominant. The quantitative dependence of $\eta_{PL}$ on $B$ depends on the spin-flip rate $\gamma_s$, exciton lifetime $\gamma_l$ and the phonon relaxation rate $\gamma_{ph}$ which appear to be comparable to each other in bilayer TMDs (see supplementary information).

In summary, we have experimentally demonstrated the Zeeman splitting in bilayer TMDCs and discussed their origin from spin-valley-layer coupling. Electrical control of orbital magnetic momentum as demonstrated previously \cite{Xu14,JonesYuEtAl2014}, together with magnetic control here will form a complete toolbox set for controlling valley and layer pseudospins. Magnetoelectric effect by the interference between electrical and magnetic field will be naturally the next step towards quantum gates or quantum entanglement between spin, valley and layer degree of freedom in bilayer platforms \cite{GongLiuEtAl2013}.  Optical stark effect by means of pseudomagnetic field has been demonstrated to control the coherence of valley pseudospins \cite{Ye2016, Schmidt2016, Wang2016}. Real magnetic control of bilayer as demonstrated here, combined with pseudomagnetic method provides access to manipulate the coherence in the bilayer system.

\subsection{Methods}

\subsubsection{Spectroscopy experiment setup}
The Raman spectra are taken at room temperature with an excitation wavelength of 532 nm using a commercial WITech confocal Raman spectrometer. We use a homemade fiber-based confocal microscope for polarization-resolved PL spectroscopy. The wavelength of the excitation is 1040 nm. The excitation power is $\sim$400$\mu$W. Polarizers and 
quarter wave planes are installed on the excitation and detection arm of the confocal microscope for polarization-selective excitation and PL detection. The PL emission is directed by an multi-mode optical fiber into a spectrometer (Princeton Instruments) with a liquid nitrogen cooled infrared camera for spectroscopy recording.   
The sample is loaded into a Quantum Design Physical Properties Measurement System (PPMS) and cooled down to liquid helium temperature.
The magnetic field is applied perpendicular to the sample plane. The magnetic field ranges from -8 T to +8 T.

\subsubsection{Preparation of MoTe2 thin flakes}

The MoTe2 single crystals were synthesized through chemical vapor transport using iodine as the transport agent. A scotch tape-based mechanical exfoliation method was used to peel thin flakes from bulk crystal onto degenerately doped silicon wafer covered with a layer of 285 nm thermally grown silicon dioxide. Optical microscopy (Olympus BX-51) was used to identify thin flake samples with different thickness via optical contrast.

%\subsubsection{Fittings/Deconvolution}

\subsection{Appendix}

\subsubsection{Experimental results for polarization-resolved photoluminescence}

\begin{figure}[h]
\centering
\includegraphics[width=0.99\linewidth]{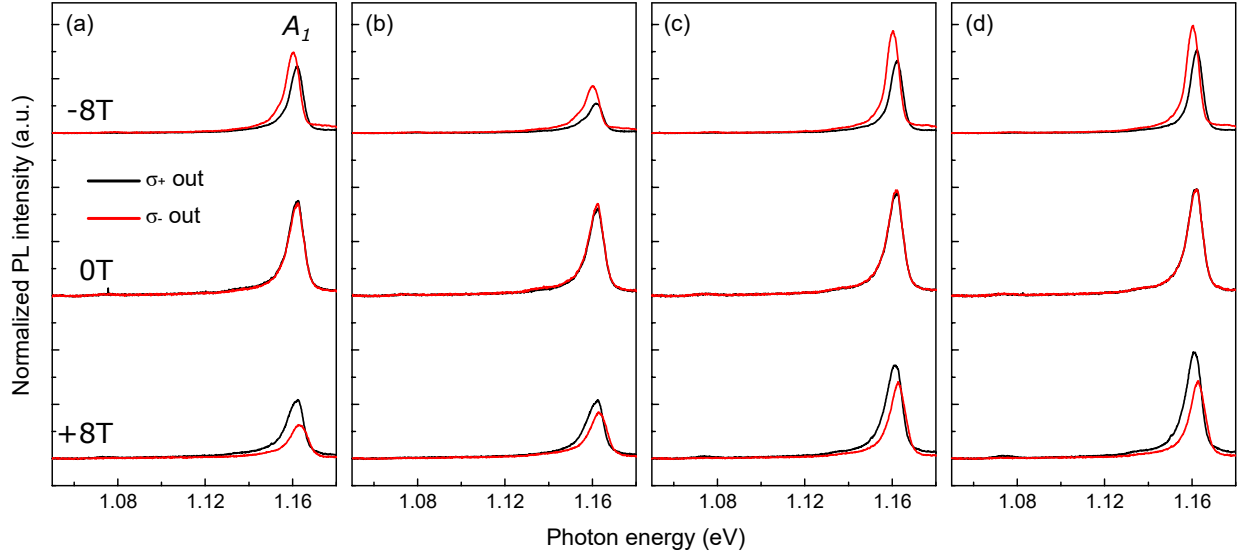}
\caption{Polarization-resolved PL
	spectra in monolayer MoTe2 at -8T, 0T and +8T. 
	The excitation polarization is set to: (a) $\sigma_+$ polarization, (b) $\sigma_-$ polarization, (c) horizontal linear polarization, (d) vertical linear polarization. The detections
	are configured to $\sigma_+$ (black solid lines) and $\sigma_-$ (red solid lines) circular polarization. $A_1$ labels the exciton peak.  }
\label{fig:fig_s1}
\end{figure}

\begin{figure}[h]
	\centering
	\includegraphics[width=0.99\linewidth]{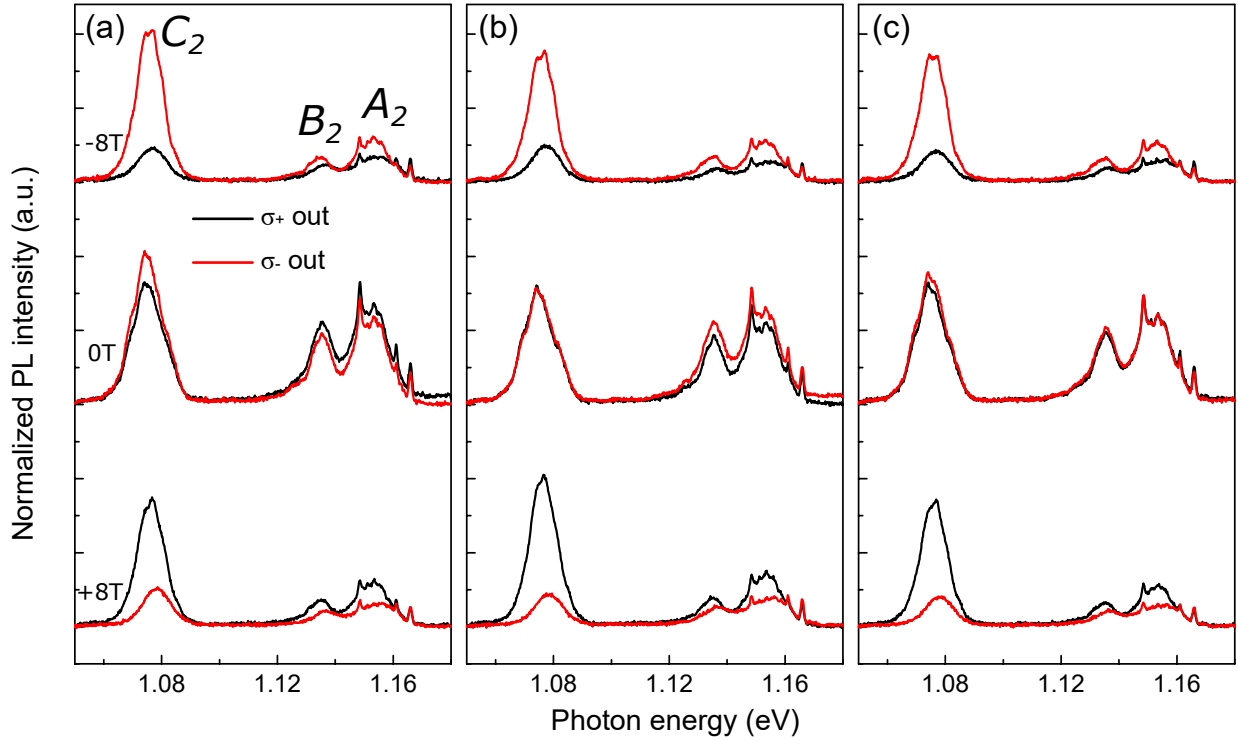}
	\caption{Polarization-resolved PL
		spectra in bilayer MoTe2 at -8T, 0T and +8T. 
			The excitation is set to: (a) $\sigma_+$ polarization, (b) $\sigma_-$ polarization, (c) horizontal linear polarization. The detections
		are configured to $\sigma_+$ (black solid lines) and $\sigma_-$ (red solid lines) circular polarization. $A_2, B_2, C_2$ represent the exciton peak, the Trion peak and the local exciton peak, respectively.}
	\label{fig:fig_s2}
\end{figure}

\begin{figure}[h]
	\centering
	\includegraphics[width=0.99\linewidth]{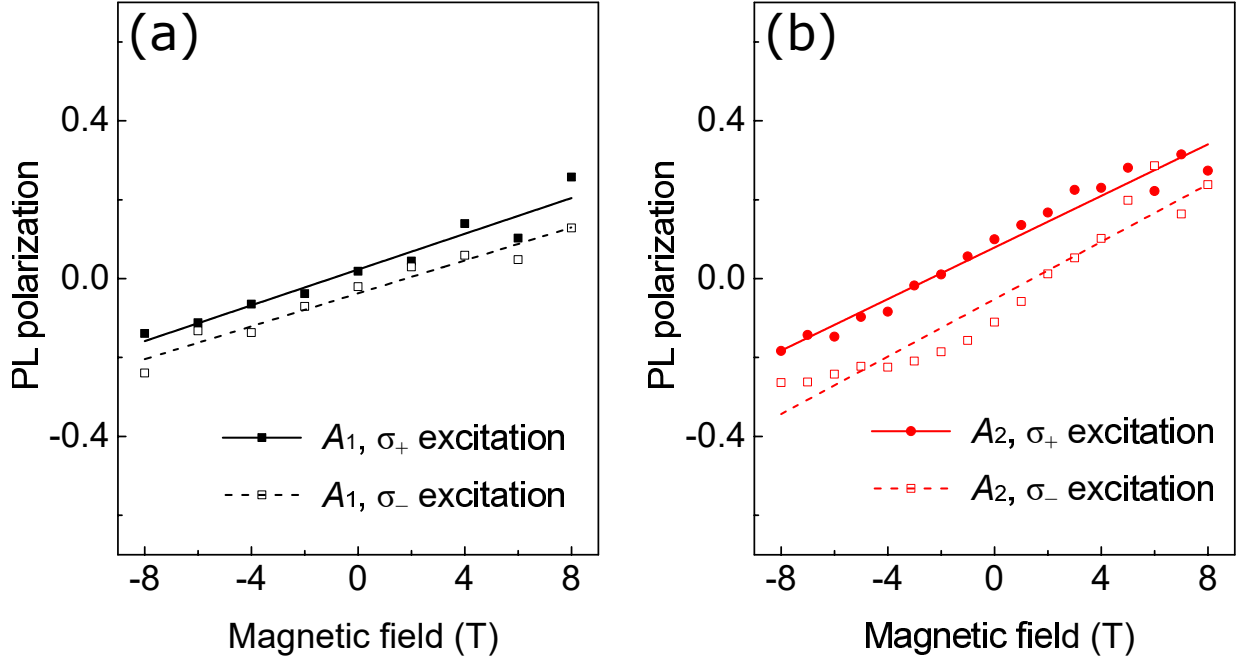}
	\caption{PL polarization vs magnetic field for exciton in monolayer ($A_1$) and bilayer ($A_2$) with different excitation polarization. (a) monolayer exciton; (b) bilayer exciton. Here the degree of the PL polarization can be defined as 
	$\eta_{PL}=\frac{I_{\sigma_+}-I_{\sigma_-}}{I_{\sigma_+}+I_{\sigma_-}}$, where $I_{\sigma_+}$ ($I_{\sigma_-}$) is the intensity of the PL emission in $\sigma_+$-out ($\sigma_-$-out) configurations.  The line shows a linear fitting of the raw data. A more detailed fitting for the output polarization can be found in Supplementary section II and Figure S5.  The solid line ($\sigma_+$ excitation) is above the dashed line ($\sigma_-$ excitation), meaning that there is positive correlation between the emission and excitation polarization.  }
	\label{fig:fig_s3}
\end{figure}

\begin{figure}[h]
	\centering
	\includegraphics[width=0.9\linewidth]{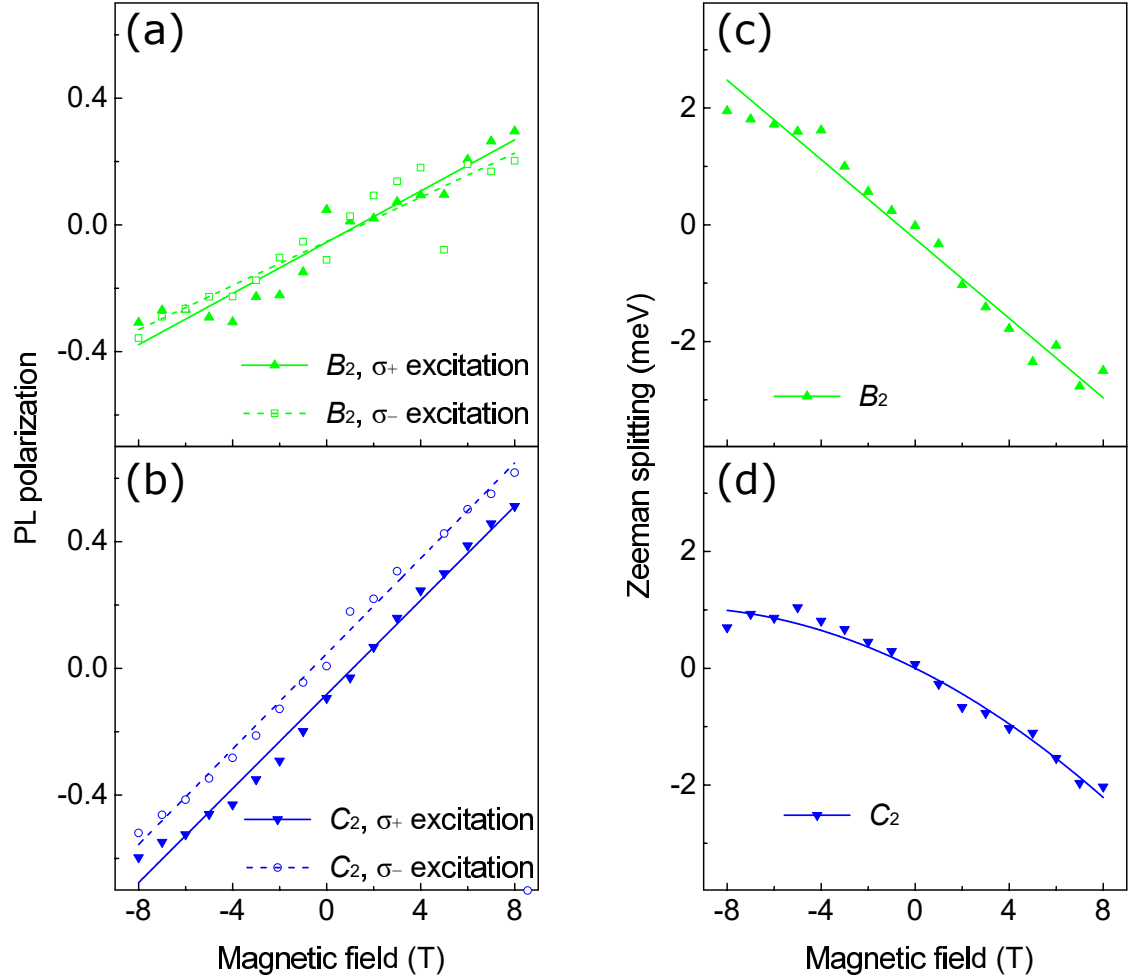}
	\caption{PL polarization and valley splitting for Trion peak and local exciton peak in bilayer. (a) PL polarization for Trion $B_2$; (b) PL polarization for local exciton peak $C_2$; (c) Zeeman splitting for Trion $B_2$; (d) Zeeman splitting for local exciton $C_2$. Local excitons has an anormous behaviour of reserved valley polarization and diamagnetic curve, which is not well understood. }
	\label{fig:fig_s4}
\end{figure}

In our experiment we pump the sample with four different polarization state: right-circular, left-circular, horizontal-linear and vertical-linear. The photoluminescence (PL) spectra are shown in Figure \ref{fig:fig_s1} for the monolayer and \ref{fig:fig_s2} for the bilayer. It can be seen that there is no dependence of excitation polarization for monolayer. For bilayer, although the emission polarization is predominantly independent of excitation polarization, there is indeed a finite correlation between the excitation and emission polarization, which we show in Fig. \ref{fig:fig_s3}. Such valley correlation has been observed in bilayer WS2 too \cite{BairenZhu14}.  Except for exciton peak $A2$ as discussed in the main text, Zeeman splitting of Trion peak $B_2$ and local exciton peak $C_2$ shows the dependence on magnetic field too (Figure. \ref{fig:fig_s4}).

%The setup is shown in Figure \ref{fig:fig_s0}. Compared with the setup in the main text, two half wave plates are added for generating and analyzing the linearly polarized light.
% We also extract the horizontal and vertical linearly polarized components of the PL emission (see Figure \ref{fig:fig_s3}). The photon energy of the linearly polarized emission lies between the $\sigma_+$ and $\sigma_-$-polarized emission. There is not magnetic field dependent shift of the linearly polarized emission.

%\begin{figure}[tbph]
%	\centering
%	\includegraphics[width=0.49\linewidth]{tu/fig_s000}
%	\caption{Polarization-resolved PL
%		spectra of the neutral exciton state taken at -8 T, 0 T and +8 T at 4 K in monolayer \ce{MoTe2}}
%	\label{fig:fig_s000}
%\end{figure}

\subsubsection{Model to explain the magnetic field dependence of PL polarization}
As discussed in the main text, we assume that interlayer coupling is suppressed due to large spin-orbit coupling. Below we use a rate equation model to estimate the decay rates. We put the rate of optical pumping to both $\sigma^+$ and $\sigma^-$ excitons is the same, $\Gamma_{pump}$. Let the energy conserving spin flip rate be $\gamma_s$, the phonon emission rate be $\gamma_e$ while the phonon absorption rate be $\gamma_a$. Assuming a lattice temperature of $T$, we have $\gamma_a = \gamma_e e^{-\Delta_B/k_BT}$. Furthermore, we assume $\gamma_l$, the total lifetime of both $\sigma^+$ and $\sigma^-$ excitons to be the same. As the spin-flip process is assumed to be energy conserving, we can work with three levels to describe the conversion process between $\sigma^+$ and $\sigma^-$ exciton. Let $n_1$ ($n_3$) be the population in the higher (lower) energy exciton and $n_2$ be population in the state obtained after elastic spin-flip from the higher energy exciton.
%that the PL polarization is independent of the excitation polarization such that 

The following set of rate equations relate the steady populations -

\begin{eqnarray}
\Gamma_{pump} - (\gamma_s + \gamma_l)n_1 + \gamma_s n_2 &=& 0 \\
\gamma_s n_1 - (\gamma_e + \gamma_s) n_2 + \gamma_a n_3 &=& 0 \\
\Gamma_{pump} + \gamma_e n_2 - (\gamma_l + \gamma_a) n_3 &=& 0.
\end{eqnarray}

\begin{figure}[tbph]
	\centering
	\includegraphics[width=0.69\linewidth]{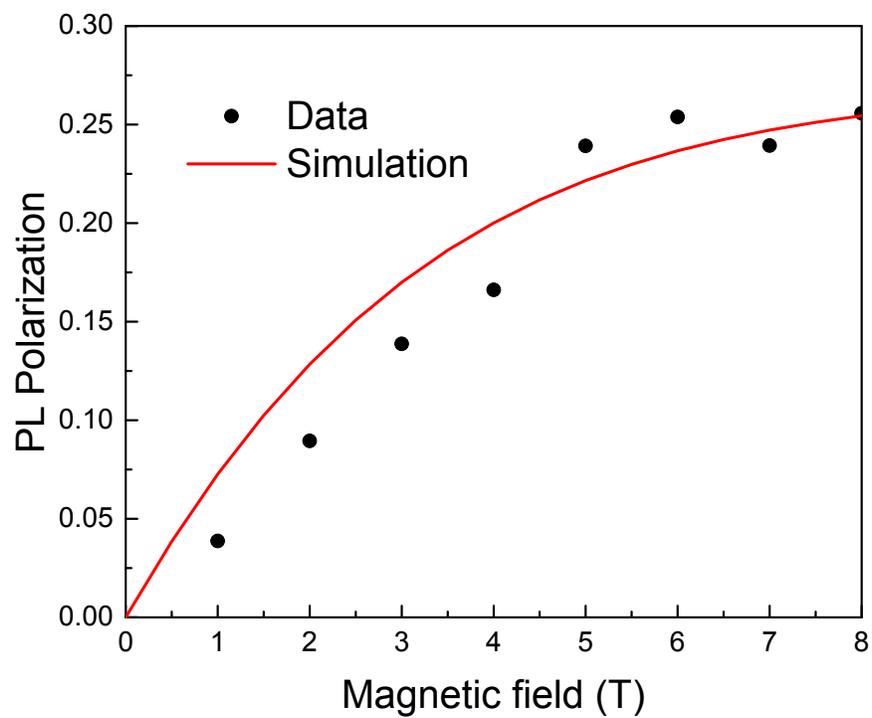}
	\caption{Fit to the PL polarization of fig.~2d of main text using the model described in the text}
	\label{fig:fig_s5}
\end{figure}

Figure \ref{fig:fig_s5} shows a fit of $\eta_{PL}$ to data in fig 2d of main text, for similar values of all rates and lattice temperature of T = 6.5 K which is slightly higher than the assumed temperature of the sample (4.2 K), possibly due to optical excitation. Our simple model qualitatively captures the magnetic field dependence of PL polarization data reasonably well. It can be concluded that all rates in the model are comparable on the scale of picoseconds.

\newpage
%
%\begin{figure}[h]
%	\centering
%	\includegraphics[width=0.39\linewidth]{fig_s0}
%	\caption{Optical setup
%		for the polarization-resolved PL spectroscopy. The optical components are:
%		achromatic lenses (AL1-3), polarizers (P1 and P2), quarter wave plates (QWP1 and QWP2),  half wave plates (HWP1 and HWP2), a 1050 nm short pass filter (SPF), a 1050 nm long pass filter (LPF) and a beam splitter (BS). The sample is placed in a helium bath cryostat with an out of plane magnetic field in a Faraday geometry (green arrow).}
%	\label{fig:fig_s0}
%\end{figure}
%
%%
%\begin{figure}[tbph]
%	\centering
%	\includegraphics[width=0.89\linewidth]{fig_s001_2}
%	\caption{(a) Polarization-resolved PL
%		spectra of the trion state taken at -8 T, 0 T and +8 T at 4 K in bilayer MoTe2. (b) Valley splitting and (c) PL polarization of the trion state emission versus magnetic field in bilayer MoTe2 at 4 K.}
%	\label{fig:fig_s001}
%\end{figure}
%
%
%\begin{figure}[tbph]
%	\centering
%	\includegraphics[width=0.89\linewidth]{fig_s002_2}
%	\caption{(a) Polarization-resolved PL
%		spectra of the local exciton state taken at -8 T, 0 T and +8 T at 4 K in bilayer MoTe2. (b) Valley splitting and (c) PL polarization of the local exciton state emission versus magnetic field in bilayer MoTe2 at 4 K. Diamagnetic shift of the valley splitting is seen.}
%	\label{fig:fig_s002}
%\end{figure}

%\begin{figure}[tbph]
%	\centering
%	\includegraphics[width=0.69\linewidth]{fig_s3}
%	\caption{PL spectra of the monolayer at +8 T. The quarter-wave plates QWP1 is set to $\sigma_+$. Right-circular (black line), left-circular (red line), horizontal-linear (green line) and vertical-linear (blue line) polarization are extracted from the PL emission.}
%	\label{fig:fig_s3}
%\end{figure}

\end{document}